\documentclass[superscriptaddress,floatfix,nofootinbib,longbibliography,prx,twocolumn]{revtex4-1}
\usepackage{graphicx,amsfonts,amssymb,amsmath,hyperref,hypcap,enumerate}
\usepackage{slashed}
\usepackage{amsmath,amssymb,bm,graphicx,dsfont}
\usepackage[latin9]{inputenc}
\usepackage{amsmath}
\usepackage{slashed}
\usepackage{dsfont}
\usepackage{amstext}
\usepackage{amssymb}
\usepackage{amsbsy}
\usepackage{amsthm}
\usepackage{epsfig}
\usepackage{graphicx}
\usepackage{bbm}
\usepackage{hyperref}
\usepackage{color}
\usepackage{slashed}
\newcommand{\be}{\begin{equation}}
\newcommand{\ba}{\begin{align}}
\newcommand{\ee}{\end{equation}}
\newcommand{\bea}{\begin{eqnarray}}
\newcommand{\eea}{\end{eqnarray}}
\newcommand{\beq}{\begin{equation}}
\newcommand{\eeq}{\end{equation}}
\newcommand{\beqn}{\begin{eqnarray}}
\newcommand{\eeqn}{\end{eqnarray}}

\newcommand{\la}{\langle}
\newcommand{\ra}{\rangle}

\newcommand{\lp}{\left(}
\newcommand{\rp}{\right)}

\renewcommand{\vec}[1]{{\bf #1}}
\renewcommand{\hat}[1]{{\widehat #1}}

\def\nn{\nonumber\\}

\begin{document}
\title{A theory of deconfined pseudo-criticality}
\author{Ruochen Ma}
\affiliation{Perimeter Institute for Theoretical Physics, Waterloo, ON N2L 2Y5, Canada}
\affiliation{Department of Physics and Astronomy, University of Waterloo, Waterloo, ON N2L 3G1, Canada}
\author{Chong Wang}
\affiliation{Perimeter Institute for Theoretical Physics, Waterloo, ON N2L 2Y5, Canada}
\date{\today}

\begin{abstract}

{
\noindent
It has been proposed that the deconfined criticality in $(2+1)d$ -- the quantum phase transition between a Neel anti-ferromagnet and a valence-bond-solid (VBS) -- may actually be pseudo-critical, in the sense that it is a weakly first-order transition with a generically long correlation length. The underlying field theory of the transition would be a slightly complex (non-unitary) fixed point as a result of fixed points annihilation. This proposal was motivated by existing numerical results from large scale Monte-Carlo simulations as well as conformal bootstrap. However, an actual theory of such complex fixed point, incorporating key features of the transition such as the emergent $SO(5)$ symmetry, is so far absent. Here we propose a Wess-Zumino-Witten (WZW) nonlinear sigma model with level $k=1$, defined in $2+\epsilon$ dimensions, with target space $S^{3+\epsilon}$ and global symmetry $SO(4+\epsilon)$. This gives a formal interpolation between the deconfined criticality at $d=3$ and the $SU(2)_1$ WZW theory at $d=2$ describing the spin-$1/2$ Heisenberg chain. The theory can be formally controlled, at least to leading order, in terms of the inverse of the WZW level $1/k$. We show that at leading order, there is a fixed point annihilation at $d^*\approx2.77$, with complex fixed points above this dimension including the physical $d=3$ case. The pseudo-critical properties such as correlation length, scaling dimensions and the drifts of scaling dimensions as the system size increases, calculated crudely to leading order, are qualitatively consistent with existing numerics.}

\end{abstract}

\maketitle



Going beyond the Landau paradigm has been a modern theme in the study of phase transitions. In the context of quantum magnetism, the prime example is the so-called deconfined quantum critical point (DQCP) -- a direct continuous transition between a Neel antiferromagnet and a valence-bond-solid (VBS) state on a square lattice\cite{deccp,deccplong}. These two states break very different symmetries (spin rotation for Neel and lattice rotation for VBS) so a direct, continuous transition is forbidden in Landau theory without further fine tuning. For $SU(N)$ spin systems with sufficiently large $N$, the existence of such non-Landau continuous transition has been firmly established both theoretically\cite{deccp,deccplong} and numerically\cite{kaulsandviklargen}, so there is no question on whether such non-Landau transition can exist. However for $SU(2)$ spins -- the most interesting case for condensed matter physicists -- the situation has been murky since the early days. 

The continuum field theory describing the DQCP, known as the (non-compact) $CP^1$ theory, is a strongly coupled gauge theory with little theoretical control. Therefore large scale numerical simulations are needed to determine whether the transition is truly continuous. Many such Monte-Carlo simulations have been carried out in the past decade, on different lattice realizations of the  DQCP\cite{SandvikJQ,melkokaulfan,lousandvikkawashima,Banerjeeetal,Sandviklogs,Kawashimadeconfinedcriticality,Jiangetal,deconfinedcriticalityflowJQ,sandvik2parameter,DCPscalingviolations,emergentso5,MotrunichVishwanath2,kuklovetalDCPSU(2),Bartosch,CharrierAletPujol, Chenetal,Aletextendeddimer,powellmonopole,AssaadDQCP,HongDQCP}, with linear system size $L$ measured in unit of lattice spacing as large as $125\sim256$ (quantum spin model\cite{Sandviklogs,sandvik2parameter}) or $640$ (classical loop model\cite{DCPscalingviolations}). Standard signatures of first-order transition (such as double-peaked probability distributions) have not been seen at the transitions in these simulations. Rather the correlation length appear to exceed the (already quite large) system size at the transition. The critical exponents extracted from finite-size scaling behaviors are roughly consistent across different simulations. However the transition does not behave like a conventional continuous transition either: the critical exponents show significant dependence on system size up to the largest size simulated. Specifically, the two exponents $\nu$ and $\eta$ drift systematically to smaller values as system size grows. Even worse, the correlation length exponent $\nu$ extracted from the largest system size ($\sim0.44$ from Refs.~\cite{DCPscalingviolations,sandvik2parameter}) is smaller than the lower bound on $\nu$ ($\sim0.511$) for a continuous transition with a single tuning parameter, found using numerical conformal bootstrap\cite{Bootstrapnu,Bootstrapreview}.
 
Another confusing issue is the emergent $SO(5)$ symmetry at the DQCP. At the Neel-VBS critical point, an emergent $SO(5)$ symmetry, rotating among the three components of Neel vector $\vec{n}$ and the real and imaginary parts of the VBS order parameter $\Phi$,  was observed numerically\cite{emergentso5}. This $SO(5)$ symmetry, absent in both the lattice models and the continuum gauge theories (such as $CP^1$), was later rationalized using dualities between different gauge theories\cite{wang2017deconfined,DualityReview} (with hints from earlier works on non-linear sigma models\cite{tanakahu,tsmpaf06}). However, assuming such an $SO(5)$ symmetry at a true critical point without further fine-tuning, the scaling dimension of the $SO(5)$ vector (in this case the Neel and VBS order parameters) is required by conformal bootstrap\cite{Bootstrapreview} to be greater than $0.76$. Numerically this scaling dimension was found to be $\sim0.62$ on the largest systems, significantly smaller than the bootstrap bound.

To resolve these discrepancies, it was proposed\cite{DCPscalingviolations,wang2017deconfined} that the DQCP for $SU(2)$ spins may actually be ``pseudo-critical". Essentially, one postulates that there is a coupling constant $\lambda$, with a flow equation under renormalization group (RG) around $\lambda=0$ given by (up to some redefinition)
\be
\label{FlowEq}
\frac{d\lambda}{dl}=\varepsilon+\lambda^2+...
\ee
 where $...$ are terms higher order in $\lambda$ and $\varepsilon$ is a small constant that is not flowing under RG. For $\varepsilon<0$, there are two fixed points: an attractive one at $\lambda_-=-\sqrt{|\varepsilon|}$, and a repulsive one at $\lambda_+=+\sqrt{|\varepsilon|}$. As $\varepsilon$ changes gradually from negative to positive, the two fixed points collide and annihilate with each other, and there is no real fixed point left. The ``pseudo-critical" scenario corresponds to a slightly positive $\varepsilon$ (ideally $0<\varepsilon\ll 1$). Some simple observations immediately follow:
 \begin{enumerate}
 \item Assuming $\lambda$ flows from $\ll-\sqrt{\varepsilon}$ to $\gg+\sqrt{\varepsilon}$. The correlation length, defined as exponential of the ``RG time" $l$ spent along the flow, is given by
 \be
 \xi=\xi_0\exp\left(\frac{\pi}{\sqrt{\varepsilon}} \right),
 \ee
 where $\xi_0$ is a non-universal constant $\sim O(1)$ depending on the UV value of $\lambda$. This can be quite large even for mildly small values of $\varepsilon$. This is sometimes also called a ``walking" coupling constant.
\item Most of the RG time is spent around $-\sqrt{\varepsilon}\lessapprox \lambda\lessapprox \sqrt{\varepsilon}$. So for small $\varepsilon$ the point $\lambda=0$ can be approximately viewed as a ``fixed point" for system size $L\ll\xi$. One can then define notions of scaling dimensions and ``relevant/irrelevant" perturbations around this pseudo-critical point. In particular, the aforementioned $SO(5)$ symmetry emerges (up to the correlation length $\xi$) if the microscopic symmetry-breaking terms are irrelevant around this fixed point $\lambda\approx 0$.

\item Even though the $\lambda\approx0$ region behaves almost like a fixed point for $L\ll\xi$, the parameter $\lambda$ is nevertheless slowly flowing. This implies that the scaling dimensions, generically as functions of $\lambda$, will be slowly drifting as the system size increases.

\item The flow equation Eq.~\eqref{FlowEq} does have two complex fixed points at $\lambda_{\pm}=\pm i\sqrt{\varepsilon}$. The pseudo-critical behavior near $\lambda=0$ on the real axis can be viewed as ultimately controlled by the complex fixed-points (even though the fixed points themselves are unreachable due to unitarity of the underlying quantum mechanical system).
 \end{enumerate}
 
The above features of the pseudo-criticality scenario could potentially resolve the existing issues in numerics. However an actual theory of the DQCP, that naturally incorporates features like pseudo-criticality and the emergent $SO(5)$ symmetry, is currently absent -- although a tentative theory for pseudo-criticality in $CP^{N-1}$ models has been qualitatively discussed in Ref.~\cite{DCPscalingviolations}. The goal of this work is to develop such a theory, and to gain a clearer picture of the origin and contents of Eq.~\eqref{FlowEq} in the DQCP. Such theories of pseudo-criticality have been developed for certain $(3+1)d$ gauge theories\cite{Gies,merge,gukov2017rg} and $(1+1)d$ $q$-state Potts models with $q>4$\cite{nienhuispotts,nauenbergscalapino,cardynauenbergscalapino,RychkovWalking1,RychkovWalking2,Mahe}.

We adopt the sigma-model approach to the DQCP. It is known that the DQCP has a ``caricature" representation in terms of a non-linear sigma model\cite{tanakahu,tsmpaf06}
\be
\label{NLsM3d}
S=\int d^3x\frac{1}{4\pi g}(\nabla\hat{N})^2+k\Gamma^{WZW}[\hat{N}],
\ee
where $\hat{N}=(n_1,n_2,n_3,{\rm Re}(\Phi),{\rm Im}(\Phi))\in S^4$ represents the combined Neel-VBS order, $g$ is the coupling strength, $\Gamma^{WZW}$ is the standard Wess-Zumino-Witten (WZW) term (well-defined since $\pi_{3+1}(S^4)=\mathbb{Z}$) with a quantized coefficient $k$, and in the case of the DQCP $k=1$. The physical significance of $\Gamma^{WZW}$ is that a vortex of the complex operator $\Phi$ traps a spin-$1/2$ moment, manifested as an effective $(0+1)d$ WZW term for $(n_1,n_2,n_3)$ -- this is exactly the feature expected for the DQCP from the lattice scale\cite{mlts04}. 

However, Eq.~\eqref{NLsM3d} is only a caricature because, as a continuum field theory, its dynamics is only well-defined in the weak-coupling regime, where the $SO(5)$ symmetry is spontaneously broken and $\la\hat{N}\ra\neq0$. Turning on a Neel-VBS anisotropy $n_1^2+n_2^2+n_3^2-|\Phi|^2$ will induce a Neel-VBS transition, but a strongly first-order one. Realizing the DQCP, even in the pseudo-criticality scenario, requires accessing some strong-coupling regime which is not well-defined on its own.

It is instructive to look at what happened in a much better understood case: the WZW sigma model at $k=1$ in $(1+1)d$, with target space $S^3$ (so the order parameter is an $SO(4)$ vector). The Lagrangian takes the same form as Eq.~\eqref{NLsM3d} except every term lives in one dimension lower and $\hat{N}\in S^3$. This theory is asymptotically free, so the free Gaussian fixed point is unstable in IR (as required also by Mermin-Wagner). The coupling strength will always flow to a critical value $g_c$ which is nothing but the famous $SU(2)_1$ CFT (recall that $SU(2)\sim S^3$)\cite{witten1994non}. This is also the theory describing the critical spin-$1/2$ Heisenberg-Bethe chain\cite{FradkinBook}, and can be viewed as the close relative of the DQCP in $(1+1)d$.

We now propose a theory of WZW non-linear sigma model, formally defined in space-time dimension $d=2+\epsilon$, with target space $S^{3+\epsilon}$ (so the symmetry is $SO(4+\epsilon)$). We do not attempt to explicitly write down the corresponding action (especially the WZW term) since we do not know how to precisely define the winding number of $S^{3+\epsilon}$ on another $S^{3+\epsilon}$. We simply postulate the existence of such theory as some kind of analytic continuation of WZW theories in general (positive integer) $d$ space-time dimensions with target space $S^{d+1}$ -- actions like Eq.~\eqref{NLsM3d} are always well-defined for these theories since $\pi_{d+1}(S^{d+1})=\mathbb{Z}$.

\begin{figure}
\begin{center}
\includegraphics[width=3.2in]{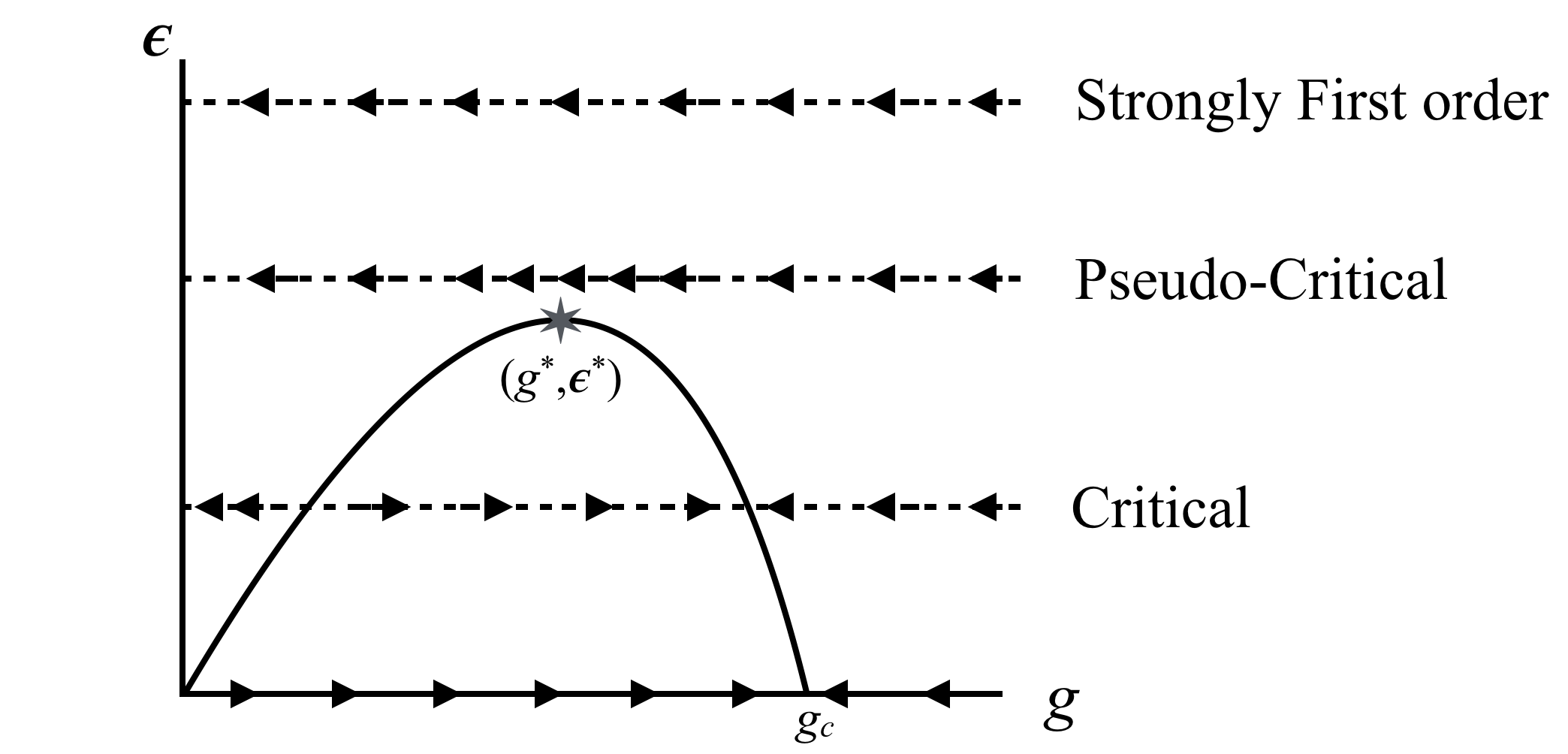}
\end{center}
\caption{Schematic RG flow of the coupling strength $g$, of the WZW sigma model in $2+\epsilon$ space-time dimensions with target manifold $S^{3+\epsilon}$, at different values of $\epsilon$. Depending on whether the physical dimension of the DQCP ($\epsilon=1$) is below, well above, or slightly above $\epsilon^*$, the system will show critical, strongly first order, or pseudo-critical behavior, respectively.
} \label{Flow}
\end{figure}

Let us first ask what are the possible scenarios based on qualitative considerations. We expect the RG flow of $g$ to look like Fig~\ref{Flow}. At $\epsilon=0$ there is a stable fixed point at $g=g_c$ and an unstable Gaussian fixed point at $g=0$. At small positive $\epsilon$, the attractive fixed point will continue in some fashion from $g_c$, but the Gaussian fixed point turns from unstable to stable because Mermin-Wagner no longer applies in dimension higher than two. Therefore another repulsive fixed point must emerge between the Gaussian ($g=0$) and the attractive one (around $g_c$). As $\epsilon$ increases, both the repulsive and attractive fixed points will continue in some fashion, but we expect them to collide and annihilate each other at some critical $\epsilon^*$ -- otherwise this would lead to interacting, non-supersymmetric CFTs in arbitrarily high dimensions, which is hard to imagine. As for the physical case of $\epsilon=1$, there are three possible scenarios: (a) $1<\epsilon^*$, and the attractive fixed point describes the truly continuous DQCP, (b) $\epsilon^*$ significantly below $1$, and the transition is strongly first order, and (c) $\epsilon^*$ slightly below $1$, and the system shows pseudo-critical behavior before eventually crossing over to first order transition at large system size. Based on existing numerical results, we expect scenario (c) to be the physical one, and the small constant in Eq.~\eqref{FlowEq} is $\varepsilon\propto (1-\epsilon^*)$.

Let us now try to be slightly more quantitative. The WZW sigma model can be perturbatively controlled if the WZW level $k$ is large. In this case $g_c(\epsilon=0)\sim1/k$, and we will see that we also have $g^*\sim\epsilon^*\sim1/k$. Of course $k=1$ for the physical case, so an expansion in $1/k$ (especially to low order) may not be trusted quantitatively. Nevertheless, just like usual small $\epsilon$ or large $N$ expansions, such a calculation can offer valuable insights, especially when combined with other approaches such as lattice simulations.

The next question is how to compute the perturbative RG equation in $2+\epsilon$ dimensions with a WZW term for $S^{3+\epsilon}$ -- after all, we do not even have a Lagragian for such theories. However we do not need to have a Lagrangian -- all we need to do is to analytically continue the perturbative RG flow equations in integer dimensions $d$ with target manifold $S^{d+1}$. The flow equations for integer $d\geq2$ take the form
\bea
\frac{dg}{dl}&=&-\epsilon g+2g^2-F(d)k^2g^{2+d}+..., \nn
\frac{dk}{dl}&=&0,
\eea
where the second equation simply comes from level quantization of WZW term, the $2g^2$ term in the first equation is a standard result for non-linear sigma model, the $k^2g^{2+d}$ term is the leading order contribution from WZW term (see Appendix A for more details) and $F(d)$ is some function of $d$. It is known\cite{witten1994non} that $F(2)=2$. We assume that the continuation of the second equation to fractional $\epsilon$ is trivially $dk/dl=0$, namely we assume that the WZW level is quantized even for fractional $\epsilon$ (something like $\pi_{3+\epsilon}(S^{3+\epsilon})=\mathbb{Z}$). Now assuming an analytic continuation of $F(d)$ exists, then for $d=2+\epsilon$ with $g\sim\epsilon\sim1/k$, the leading order flow equation simply becomes
\be
\label{beta}
\frac{dg}{dl}=-\epsilon g+2g^2-2k^2g^4+...
\ee
In particular, we only need the zeroth order value of the $F(2+\epsilon)$ term -- the calculation would otherwise be much more complicated. 

The fixed points from Eq.~\eqref{beta} are given by
\be
\frac{\epsilon}{2}=g-k^2g^3,
\ee
which indeed behave as Fig.~\ref{Flow}. The critical dimension and coupling strength are given, to leading order in $1/k$, by
\bea
\epsilon^*&=&\frac{4}{3\sqrt{3}k}\approx \frac{0.77}{k}, \nn
g^*&=&\frac{1}{\sqrt{3}k}.
\eea

Now consider the theory just above the critical dimension, $\epsilon=(1+\alpha)\epsilon^*$ with $0<\alpha\ll1$. Eq.~\eqref{beta} then reduces to Eq.~\eqref{FlowEq} to leading order in $\alpha$, with $\lambda=2(g-g^*)$. The correlation length is now (again to leading order in both $1/k$ and $\alpha$)
\be
\label{xi}
\xi=\xi_0\exp\lp\frac{\pi}{\sqrt{2\alpha\epsilon^*g^*}}\rp=\xi_0\exp\lp\frac{3\pi k}{\sqrt{8\alpha}} \rp.
\ee

Putting $k=1$ into the above results, we get $\epsilon^*\approx0.77$. The physical case of $\epsilon=1$ corresponds to $\alpha\approx0.3$, which then gives the estimated correlation length $\xi\approx 440\xi_0$. These are indeed consistent with pseudo-criticality! This is also qualitatively consistent with existing numerics, in the sense that it can be easily larger (but not too much larger) than the simulated system size. 

We can also estimate critical exponents at the deconfined pseudo-critical point to leading order. The scaling dimensions of rank-$l$ (symmetric traceless) tensors of the $SO(4+\epsilon)$ group are given by
\be
\label{exponents}
\Delta_l=\frac{l(l+2)}{2}g^*=\frac{l(l+2)}{2\sqrt{3}k},
\ee
where the first identity comes from standard non-linear sigma model calculations without the WZW term -- the WZW only affects the result through $g^*$ (see Appendix A). At $k=1$, this gives $\Delta[l=1]\approx0.87$ and $\Delta[l=2]\approx 2.3$. For $l=1$ (Neel/VBS order parameter) the numerical simulations give $\Delta[l=1]_{Num}=(1+\eta)/2\approx 0.62\pm 0.1$, while for $l=2$ (Neel-VBS anisotropy) the numerical value is roughly $\Delta[l=2]_{Num}=3-1/\nu\approx 1.0\pm0.3$. The error bar comes from sampling different works, on difference system sizes with different schemes used to extract the exponents. Our estimated value (in $1/k$) for the vector order parameter is in qualitative agreement with the numerical values. In fact the estimation is far better than a similar $O(1/k)$ estimation in $2d$, where the exact result is known to be $\Delta[l=1]_{2d}=3/2(k+2)=1/2$ while the $O(1/k)$ estimation gives $3/2$. In some sense this means that theories at $\epsilon>0$ are less strongly coupled than the $2d$ $SU(2)_1$ theory so perturbative calculations become more reliable. Our estimation for the rank-$2$ tensor is less impressive -- this is perhaps not too surprising since a similar estimation in $2d$ gives even larger error than the vector case. Furthermore, an estimation of rank-$4$ tensor shows that it is strongly irrelevant -- this is crucial for the emergence of $SO(5)$ at the DQCP since, in the context of DQCP, rank-$4$ tensors are allowed by microscopic symmetries as perturbations\cite{deccp,deccplong}.

Eq.~\eqref{exponents} also implies that the scaling dimensions will drift downward as the system size grows, since $g$ flows slowly to smaller and smaller values. This feature is also in agreement with numerical results. We can estimate the amount of drift at $O(1/k,\alpha)$. Assuming at system size $L_0$ the coupling constant reaches $g^*$, then for $L$ not too far away from $L_0$ (specifically $|\ln (L/L_0)|\ll \ln(\xi/\xi_0)$), the relative drift in $\Delta_l$ is roughly (see Appendix A for more details)
\be
   \frac{\Delta_l(L)-\Delta_l(L_0)}{\Delta_l(L_0)}\approx-0.23\ln (L/L_0),
\ee
which appears to be qualitatively consistent with the numerically observed drifts for the correlation length exponent $\nu$\cite{sandvik2parameter,DCPscalingviolations}.

We can also consider the $k=2$ case. Repeat the analysis above one obtains $\xi(k=2)\sim 190\xi_0$, which means a weaker but potentially observable pseudo-critical behavior -- the actual number is less reliable since $\epsilon^*(k=2)\sim0.38$ is further away from the physical dimension, and therefore the small $\alpha$ expansion is not justified. Note that since we expect operator scaling dimensions to reduce as $k$ becomes larger, the $k=2$ theory may have additional relevant operators such as the rank$-4$ tensors. The $k=2$ theory may potentially describe the Neel-columnar VBS transition of spin$-1$ anti-ferromagnets on square lattice\cite{wang2015nematicity}. However further fine-tuning will be required, which makes the theory multi-critical, if the rank$-4$ tensors are relevant -- this is consistent with recent numerics on spin$-1$ systems on square lattice\cite{KaulS1} in which a strong first-order transition was observed.

In summary, we have proposed a WZW non-linear sigma model in $(2+\epsilon)$ space-time dimension, with target space $S^{3+\epsilon}$ and global symmetry $SO(4+\epsilon)$, as an interpolation between the $SU(2)_1$ WZW CFT in $2d$ and the DQCP in $3d$. We argued on general ground that a fixed-point-annihilation should happen at some finite $\epsilon^*$, above which there is no real fixed point. We then argued, based on a crude $O(1/k)$ estimation and its consistency with existing numerics, that $\epsilon^*$ is slightly smaller than the physical value $\epsilon=1$ for the DQCP. Therefore the DQCP shows pseudo-critical behavior before eventually crossing over to a first-order transition as the system size exceeds the large correlation length. The pseudo-critical properties, calculated crudely in $O(1/k)$, are in qualitative agreement with existing numerics. We emphasize that, just like many other calculations in critical phenomena like $O(\epsilon)$ or $O(1/N)$, our $O(1/k)$ calculation is by no means a proof of pseudo-criticality in the DQCP since in reality $k=1$. Rather it gives a scenario, or a picture, that potentially describes the correct physics and is broadly consistent with existing numerics.

There are many possible future directions following our work. The most obvious one is to try to give the $S^{3+\epsilon}$ WZW theory an intrinsic definition, instead of simply assuming that a reasonable analytic continuation from integer dimensions exists (as we did here). More practically, how do we compute the perturbative RG flow equation beyond leading order? Another open problem is to extend the pseudo-critical theory to the easy-plane DQCP (which received stronger numerical support of the pseudo-critical scenario recently\cite{Zhao2018,Nahum2018}). Yet another question is how one could further generalize such theories, for example to other types of target space beyond spheres. Specifically, can we find another type of target space that pushes $\epsilon^*$ well above $1$, so that a true critical point of this type appears in $(2+1)d$? Can we even push it far enough to have a non-trivial fixed point in $(3+1)d$? These are all open questions to be explored in the future.

\begin{figure}[h]
\begin{center}
\includegraphics[width=2.4in]{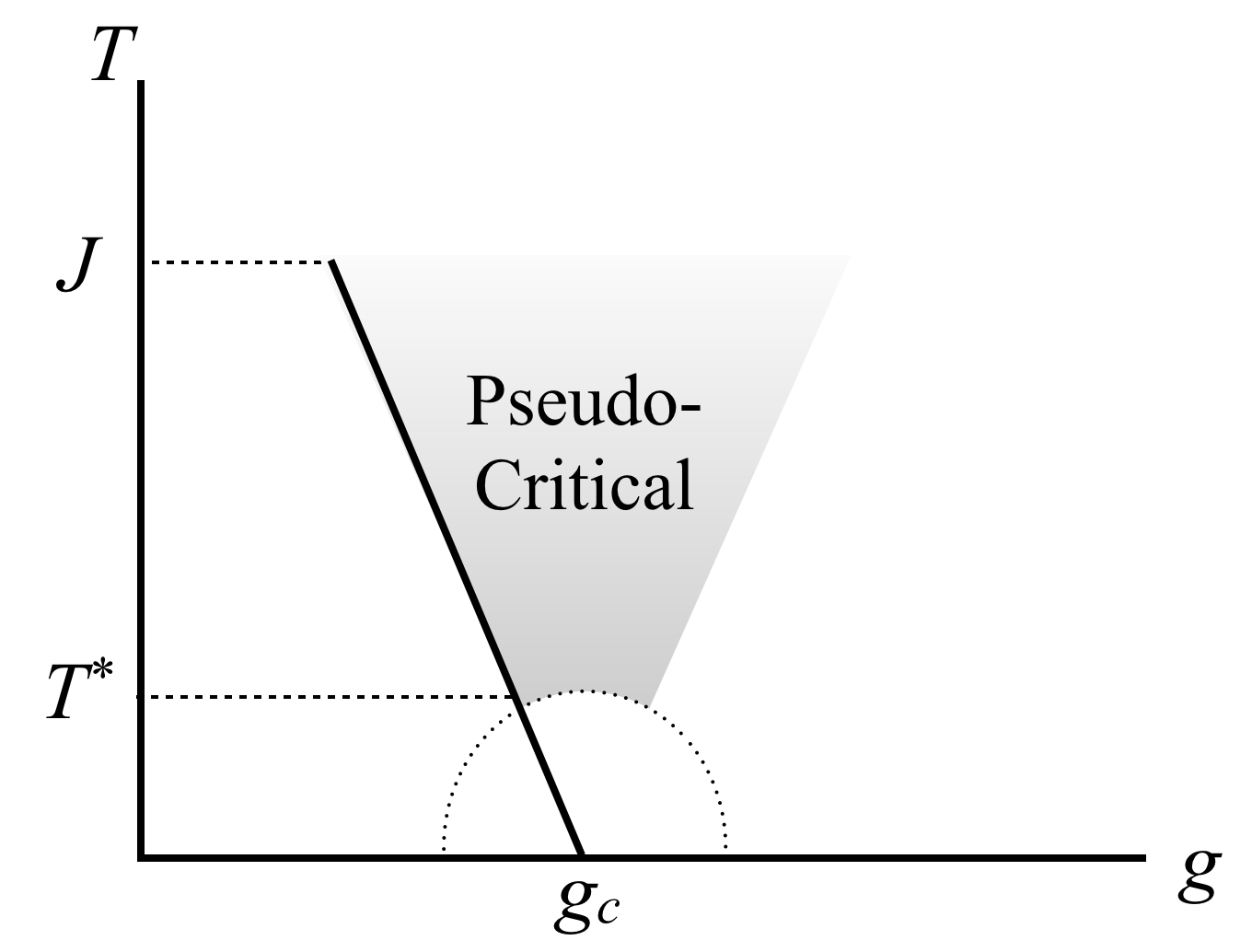}
\end{center}
\caption{Schematic phase diagram of pseudo-criticality at finite temperature. The classic ``critical fan" appears as long as the temperature is well below the microscopic energy scale $J$ and well above a very low cross-over temperature $T^*\sim J\exp(\pi/\sqrt{\varepsilon})$. Below $T^*$ the system crosses over to a conventional first order transition. 
} \label{PD}
\end{figure}

We end by emphasizing that pseudo-criticality is particularly interesting for quantum phase transitions: at finite temperature, the classic ``critical fan" appears as long as the temperature is well below the microscopic energy scale $J$ and well above a very low cross-over temperature $T^*\sim J\exp(\pi/\sqrt{\varepsilon})$. Below $T^*$ the system crosses over to a conventional first order transition. The schematic phase diagrams is shown in Fig.~\ref{PD}.

{\emph{Note added}:} During the completion of this manuscript, we became aware of an independent work by Adam Nahum which overlaps significantly with ours. 

{\textbf{Acknowledgements}}: We thank Yin-Chen He, Adam Nahum, and Djordje Radicevic for illuminating discussions. We thank Adam Nahum for sharing results prior to publication. Research at Perimeter Institute is supported by the Government of Canada through the Department of Innovation, Science and Economic Development Canada and by the Province of Ontario through the Ministry of Research, Innovation and Science.



\appendix

\section{Renormalization group calculation for WZW NL$\sigma$M}
In $d$ dimensional spacetime, we begin with an $SO(d+2)$ NL$\sigma$M with a Wess-Zumino-Witten (WZW) term at level $k$, which is in general well-defined as a continuum quantum field theory in the low temperature (ordered) phase:
\begin{widetext}
\begin{equation}
    S=S_{SO(d+2)}+ ik\Gamma[n_a]=\frac{1}{4\pi g}\int d^dx(\partial n_a)^2
    +i\frac{2\pi k}{(d+1)!\Omega_{d+1}}\int d^{d+1}y\,\epsilon^{\mu_1\mu_2...\mu_{d+1}}\epsilon^{aa_1...a_{d+1}}n_a\partial_{\mu_1}n_{a_1}...\partial_{\mu_{d+1}}n_{a_{d+1}},
\end{equation}
\end{widetext}
in which $n_a$ $(a=1,...,d+2)$ is a real $(d+2)$-component unit vector. The field $n_a$ defines a map from spacetime $S^d$ to the target space $S^{d+1}$, and $\Gamma[n_a]$ is the ratio of the surface area in $S^{d+1}$ traced out by $n_a$ to the total area of $S^{d+1}$, $\Omega_{d+1}$. $n_a(y)=n_a(x,u)$ is a smooth extension of $n_a(x)$ such that $n_a(x,0)=(0,...,1)$ and $n_a(x,1)=n_a(x)$. $\Gamma$ is well-defined modulo $\Gamma\to\Gamma+2\pi$ due to the existence of topologically inequivalent extensions, which are classified by $\pi_{d+1}(S^{d+1})\simeq Z$. $S_{SO(d+2)}$ is the non-topological part of the NL$\sigma$M action. In addition to the $SO(d+2)$ symmetric kinetic term, it may also contains anisotropy terms, which are assumed to be irrelevant in the pseudocritical regime \cite{wang2017deconfined}. By power counting, the theory is renormalizable in two spacetime dimension but nonrenormalizable for $d>2$. However, it may formally be defined in a double expansion in powers of the coupling $g$ and $\epsilon=d-2$ \cite{brezin1976spontaneous,brezin1976renormalization}. As we will see shortly, this expansion can be perturbatively controlled in the large $k$ limit. We follow the choice of parametrization of $S^{d+1}$ in Ref.~\cite{cardy1996scaling}, since it enables us to calculate the beta function and anomalous dimension of soft operators separately:
\begin{align}
    n_i=&t_i\,\,\,\,\,(1\leq i\leq d)\\
    n_{d+1}=&\sqrt{1-\mathbf{t}^2}\mathrm{cos} \theta\\
    n_{d+2}=&\sqrt{1-\mathbf{t}^2}\mathrm{sin}\theta.
\end{align}
One can easily see that the functional measure changes simply as 
\begin{equation}
    \prod_a d n_a(x)\delta(\mathbf{n}^2-1)=\frac{1}{2}d\theta(x)\prod_i d t_i(x),
\end{equation}
hence we have no additional contribution from the Jacobian. The non-topological part of the action becomes, in terms of renormalized fields,
\begin{equation}
\begin{split}
\label{eq:kinetic}
    S_{SO(d+2)}&=\frac{1}{4\pi Z_gg\mu^{-\epsilon}}\int d^dx[(1-Z_tt^2)(\partial_\mu\theta)^2\\
    &+(\partial_\mu\sqrt{1-Z_t\mathbf{t}^2})^2+Z_t(\partial_\mu \mathbf{t})^2],
\end{split}
\end{equation}
in which $\mu$ is a parameter with dimension of mass. Note that the $\theta$ variable has no field strength renormalization since it is a phase angle with definite periodicity of $2\pi$, analogous to the case in $2d$ XY model. At zeroth order the WZW term does not renormalize the kinetic terms of $t$ and $\theta$ due to the $\epsilon$ tensor. In order for a perturbative expansion, one needs to identify the vertex that contributes to the lowest order in $g$, from the expansion of WZW term. This is the vertex that contains $d$ $t_i$ fields and one $\theta$ field. Note that 
\begin{equation}
    n_{d+1}\partial_\mu n_{d+2}-n_{d+2}\partial_\mu n_{d+1}=\partial_\mu\theta+...
\end{equation}
up to terms of higher order in $t$ fields. Therefore the vertex that contributes at the leading order in $g$ is 
\begin{widetext}
\begin{align}
    &\Sigma_r \frac{2\pi ki}{(d+1)!\Omega_{d+1}} \int d^{d+1}y\, \epsilon^{\mu_1...\mu_{d+1}}\epsilon^{(d+1)a_1...(a_r=d+2)...a_{d+1}}\partial_{\mu_r}\theta\partial_{\mu_1}t_{a_1}...\widehat{\partial_{\mu_r}t_{a_r}}...\partial_{\mu_{d+1}}t_{a_{d+1}}\\
    =&\Sigma_r \frac{2\pi ki}{(d+1)!\Omega_{d+1}} \int d^{d+1}y\, \epsilon^{\mu\mu_1...\mu_{d}}\epsilon^{(d+1)(d+2)a_1...a_{d}}\partial_{\mu}\theta\partial_{\mu_1}t_{a_1}...\partial_{\mu_{d}}t_{a_{d}}\\
    =&\frac{2\pi ki}{d!\Omega_{d+1}} \int d^{d+1}y\, \epsilon^{\mu\mu_1...\mu_{d}}\epsilon^{a_1...a_{d}}\partial_{\mu}\theta\partial_{\mu_1}t_{a_1}...\partial_{\mu_{d}}t_{a_{d}},
\end{align} 
\end{widetext}
in which $\widehat{\partial_{\mu_r}t_{a_r}}$ means this term is omitted. As part of the WZW action, this term is in fact a total derivative. We assume that the manifold parametrized by coordinate $y$  has a boundary on a constant-($y_{d+1}$) plane, which is our original system:
\begin{equation}
\begin{split}
    n(x,y_{d+1}=0)=&(0,0,...,1),\\
    n(x,y_{d+1}=1)=&n(x).
\end{split}
\end{equation}
The vertex then takes the form of
\begin{align}
    &\frac{2\pi ki}{d!\Omega_{d+1}} \int d^{d}x (-)\epsilon^{(d+1)\mu\mu_2...\mu_{d}}\epsilon^{a_1...a_{d}}\partial_{\mu}\theta t_{a_1}\partial_{\mu_2}t_{a_2}...\partial_{\mu_{d}}t_{a_{d}}\\
    =&(-)^{d+1}\frac{2\pi ki}{d!\Omega_{d+1}} \int d^{d}x \epsilon^{\mu_1\mu_2...\mu_{d}}\epsilon^{a_1...a_{d}}\partial_{\mu_1}\theta t_{a_1}\partial_{\mu_2}t_{a_2}...\partial_{\mu_{d}}t_{a_{d}}
\end{align}

\begin{figure}
\centering
  \includegraphics[width=0.35\textwidth]{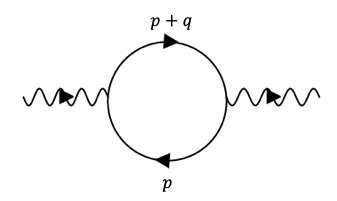} 

  \caption{The one loop correction to self-energy of $\theta$ arising from the WZW term in $2d$. The solid lines represent propagator of $t$ field while the wavy line is propagator of $\theta$.}
  \label{fig:thetaselfenergy}

\end{figure}
The beta function can be obtained by calculating self-energy of $\theta$ field. The coupling $g$ will be renormalized at leading order by a two-vertex diagram Fig.\ref{fig:thetaselfenergy}. There are $d$ internal propagators, depending on the spacetime dimension. For our purpose, it is sufficient to calculate this diagram in two dimension (zeroth order), if we assume that an analytic continuation exists, and the correction in $d=(2+\epsilon)$ is of higher order in $g$ and $\epsilon$. For a massless theory like NL$\sigma$M, near two dimension one should take special care to separate infrared and ultraviolet divergences \cite{bardeen1976phase}, and only the logarithmic divergence from UV contributes to renormalization. Physically, the IR divergences come from the absence of spontaneous breaking of continuous symmetry in two dimension. The contribution of this diagram to self-energy of $\theta$ field is

\begin{align}
    \Pi_\theta(q^2)=&(2\pi g)^2(\frac{2\pi ki}{2\Omega_{3}})^2\epsilon^{\mu_1\mu_2}\epsilon^{\nu_1\nu_2}\epsilon^{a_1a_2}\epsilon^{b_1b_2}\times\nonumber  \\
    &\int_{\Lambda/b}^\Lambda \frac{d^2p}{(2\pi)^2}[\frac{p_{\mu_1}p_{\nu_1}(p+q)_{\mu_2}(p+q)_{\nu_2}}{(p+q)^2p^2}\delta_{a_1b_1}\delta_{a_2b_2}\nonumber \\
    &+\frac{p_{\mu_1}p_{\nu_2}(p+q)_{\mu_2}(p+q)_{\nu_1}}{(p+q)^2p^2}\delta_{a_1b_2}\delta_{a_2b_1}]\\
    \sim&-\frac{2k^2g^2}{2\pi}q^2\mathrm{ln}b.
\end{align}
 In last line we keep only the divergent term, and the integral is taken over a momentum shell $|p|\in(\Lambda/b,\Lambda)$, where $\Lambda$ is a momentum cutoff. This divergence should be canceled by the coupling constant renormalization,
 \begin{equation}
     -\frac{2k^2g^2}{2\pi}q^2\mathrm{ln}b-\frac{1}{2\pi g}(\frac{1}{Z_g}-1)q^2=0.
     \label{eq:couplingre}
 \end{equation}
 The renormalization group equation follows from the invariance of the bare coupling under a change of the rescaling parameter $b\sim e^l$:
 \begin{equation}
     \frac{dg}{dl}=-g\frac{d\mathrm{ln}Z_g}{dl}=-2k^2g^4
 \end{equation}
 up to terms of higher order in $g$. Combining with contributions that arise from the loop expansion of interactions in the non-topological part $S_{SO(d+2)}$ (which is the ordinary NL$\sigma$M), one can obtain the beta function (note that here we have $n=d+2=4+\epsilon$)
\begin{equation}
\begin{split}
    \beta(g)=\frac{dg}{dl}=&-\epsilon g+(2+\epsilon)g^2+(2+\epsilon)g^3-(2k^2+A)g^4\\
    &+O(g^5,\epsilon g^4,...),
\end{split}
\end{equation}
where $A$ is a quadratic polynomial of $n$ while independent of $k$, determined by three loop calculations \cite{hikami1978three}. Note that at order $O(g^4)$ and below, the above two parts contribute additively. Take the large $k$ limit (which means $A$ can be ignored), in $(2+\epsilon)$ dimension the consistent scaling is to take $g$ to be $O(\frac{1}{k})$ as at the non-trivial fixed point of WZW model in two dimension \cite{witten1994non}, and $\epsilon$ to be $O(\frac{1}{k})$. The $g^3$, $\epsilon g^2$ terms in the beta function can therefore be simply ignored, and the beta function and fixed point equation become 
\begin{align}
\beta(g)&=-\epsilon g+2g^2-2k^2g^4
\label{eq:betafunction}
\\
    \epsilon&= 2g-2k^2g^3.
\end{align}
It is easy to obtain $g^*=\frac{1}{\sqrt{3}k}$, $\epsilon^*=4/(3\sqrt{3}k)$. Take $k=1$, $\epsilon^*\sim0.77$. We can also estimate the correlation length in the pesudocritical regime in dimension $d=2+\epsilon\gtrsim 2+\epsilon^*$. Parametrize $\epsilon=(1+\alpha)\epsilon^*$ and $g=g^*+\delta g$, and expand the beta function in powers of $\alpha$ and $\delta g$, 
\begin{equation}
    \frac{d\delta g}{dl}\sim -2\delta g^2-\frac{4}{3^\frac{3}{2}k}\alpha\delta g-\frac{4\alpha}{9k^2}+O(\delta g^3).
\end{equation} 
Using the freedom to make redefinations of the coupling $\delta g$ gives 
\begin{equation}
    \frac{d\lambda}{dl}=-\lambda^2-\frac{8\alpha}{9k^2}+O(\alpha^2),
    \label{flowequation2}
\end{equation}
where $\lambda=2(\delta g+\alpha/(3^\frac{3}{2}k
))$. For $\epsilon\gtrsim\epsilon^*$ the RG flows become very slow close to $\lambda=0$, and the long RG time required to traverse the pesudocritical regime generates a large correlation length
\begin{equation}
\label{length}
    \xi\sim \xi_0\mathrm{exp}(3k\pi/\sqrt{8\alpha})\sim440\xi_0,
\end{equation}
in which we set $\epsilon=1$ and $k=1$. 
\par

\begin{figure}
\centering
  \includegraphics[width=0.4\textwidth]{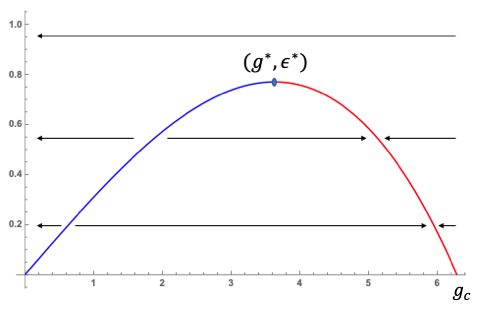} 

  \caption{RG fixed points and flows in the $(g,\epsilon)$ plane ($k=1$), showing stable (red curve) and unstable (blue curve) fixed points merging at $(g^*,\epsilon^*)$.}
  \label{fig:betafunction}

\end{figure}

\begin{figure}
\centering
  \includegraphics[width=0.35\textwidth]{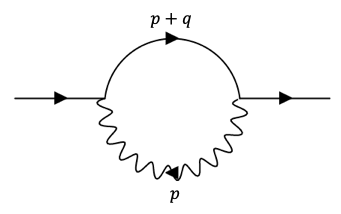} 

  \caption{The one loop correction to self-energy of $t$ field arising from the WZW term in $2d$.}
  \label{fig:tselfenergy}

\end{figure}

The renormalization of scaling operators can also be calculated by this expansion technique. In general, due to the Ward-Takahashi identity, a set of operators which correspond to the basis of an irreducible representation of the $SO(d+2)$ symmetry will not be mixed with other operators under renormalization, and there is only one renormalization constant for a given irreducible representation \cite{brezin1976renormalization,zinn1996quantum}. The anomalous dimension of the $SO(5)$ vector operator can be obtained by calculating the field strength renormalization of the $t$ field. Similarly, the loop corrections from WZW term can be calculated at $d=2$, up to terms higher order in $g$ and $\epsilon$. At one loop level, self-energy of the $t$ field acquires a divergent contribution from the WZW term, shown in Fig.\ref{fig:tselfenergy}. 

\begin{align}
 \Pi_{t}^{ab}(q^2)&=(\frac{2\pi ki}{2\Omega_3})^2\times4(2\pi g)^2\epsilon^{aa_2}\epsilon^{bb_2}\delta_{a_2b_2}\times\nonumber\\
 &\int_{\Lambda/b}^\Lambda\frac{d^2p}{(2\pi)^2}\frac{(p+q)_{\mu_1}(p+q)_{\nu_1}p_{\mu_2}p_{\nu_2}}{p^2(p+q)^2}\epsilon^{\mu_1\mu_2}\epsilon^{\nu_1\nu_2}\\
 &=-(2gk)^2\delta^{ab}\int_{\Lambda/b}^\Lambda\frac{d^2p}{(2\pi)^2}\frac{q_{\mu_1}q_{\nu_1}p_{\mu_2}p_{\nu_2}}{p^2(p+q)^2}\epsilon^{\mu_1\mu_2}\epsilon^{\nu_1\nu_2}
  \end{align}
where $a,\,b$ label the components of the vector $t_a$. The above integral can be organized in powers of the external momentum $q^2$, in which the divergent coefficient of the $q^2$ term should be removed by a counterterm related to field strength renormalization of $t$ 
\begin{equation}
    -\frac{1}{2\pi g}(\frac{Z_t}{Z_g}-1)q^2-\frac{2g^2k^2q^2}{2\pi}\mathrm{ln}b=0.
\end{equation}
Combining with our calculation of the coupling constant renormalization $Z_g$ in Eq.(\ref{eq:couplingre}), one can easily read out 
\begin{equation}
    Z_t=1+O(g^4),
\end{equation}
which means the WZW term gives no correction to scaling dimension of the vector operator, at order $O(\frac{1}{k})$ in our large $k$ expansion. The renormalization group calculations for $O(n)$ NL$\sigma$M in $2+\epsilon$ dimension in Ref.\cite{brezin1976renormalization,brezin1976anomalous} showed that the anomalous dimension of $t$ field is
\begin{equation}
    \gamma_t=\frac{n-1}{2}g+O(g^3)= \frac{3}{2}g+O(\epsilon g).
\end{equation}
If one take $n=4$, $k=1$ and $g=g^*=1/\sqrt{3}$, $\gamma_t^*\sim0.87$. 

\par
Note that if we instead restrict ourselves in two dimension, at large $k$ limit we can recover the result from $2d$ CFT with $SO(4)$ global symmetry. For $\epsilon=0$, from Eq.(\ref{eq:betafunction}) one sees that there is a critical fixed point at $g_c=1/k$, where the scaling dimension of operators in vector representation of $SO(4)$ is $3/2k$. An SO(4) vector can be realized as a $(\frac{1}{2},\frac{1}{2})$ representation of $2d$ $SU(2)\otimes SU(2)$ CFT, which has conformal weight \cite{ginsparg1988applied} 
\begin{equation}
   \Delta=\frac{3}{2(k+2)},
\end{equation}
consistent with our result at large $k$ limit.
\par
\begin{figure}
\centering
  \includegraphics[width=0.35\textwidth]{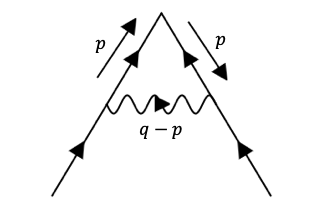} 

  \caption{The renormalization of local operator $\phi^2$}
  \label{fig:rank2}

\end{figure}

Calculation of the scaling dimension of the rank two tensor is also straightforward. We can add one such operator to the action and check how it interacts with other vertices. In our parametrization only an $SO(2)_\theta\times SO(d)_t$ subgroup is manifest. A rank two tensor of $SO(d+2)$ can be projected onto irreducible representations of $SO(2)_\theta\times SO(d)_t$. Specifically, we can calculate the scaling of the rank 2 operator of $SO(d)_t$, and it will not be mixed with other representations under renormalization. Following our assumption that the result in $2+\epsilon$ dimension can be seen as an analytic continuation from $2d$, where the correction is of higher order of $\epsilon$ and $g$, we simply do the loop expansion in $2d$. This can be easily done by a change of variable
\begin{align}
    t_1+it_2&=\sqrt{2}\phi\\
    t_1-it_2&=\sqrt{2}\phi^*.
\end{align}
And the Lagrangian of $O(4)$ NL$\sigma$M in $2d$ reads
\begin{equation}
    \begin{split}
        S=&\frac{1}{4\pi g}\int d^2x[(1-2|\phi|^2)(\nabla\theta)^2\\
        &+(\nabla\sqrt{1-2|\phi|^2})^2+2|\nabla\phi|^2].
    \end{split}
\end{equation}
The vertex takes the form of
\begin{equation}
    -\frac{2\pi ki}{2\Omega_{3}} \int d^{2}x \epsilon^{\mu_1\mu_2}\epsilon^{a_1a_2}\partial_{\mu_1}\theta \phi_{a_1}\partial_{\mu_2}\phi_{a_2},
\end{equation}
where $\phi_1=\phi$, $\phi_2=\phi^*$ and we have now $\epsilon^{12}=-\epsilon^{21}=i$, $\epsilon^{11}=\epsilon^{22}=0$. We add to the action a charge two operator of the $SO(2)_t$ symmetry, 
\begin{equation}
    {u\mu^2}\int d^2x \phi^2=\frac{u\mu^2}{2}\int d^2x (t_1^2-t_2^2+2it_1t_2).
\end{equation}
Here $u$ is a dimensionless coupling constant. The loop diagram which contributes to renormalization of $u$ is shown in Fig.\ref{fig:rank2},
\begin{equation}
\begin{split}
    -{4u\mu^2}(\frac{2\pi ki}{2\Omega_3})^2&(2\pi g)^3\epsilon^{\mu_1\mu_2}\epsilon^{\nu_1\nu_2}\\
    &\int_{\Lambda/b}^\Lambda\frac{d^2p}{(2\pi)^2}\frac{p_{\mu_2}p_{\nu_2}(q-p)_{\mu_1}(q-p)_{\nu_1}}{(p^2)^2(q-p)^2},
    \end{split}
\end{equation}
where $q$ is an external momentum. In other word, we define the normalization condition of local operator $\phi^2$ based on a Green's function $\sim\langle\phi^*\phi^*\phi^2\rangle$. The $q$-independent part, which is the only ultraviolet divergent part of the integral, should be removed by a counterterm $\sim-(Z_u-1)u\mu^2$. One can easily see that this part vanishes due to the $\epsilon$ tensors. Therefore the WZW term affects the scaling dimension only through $g^*$. We can find from Ref.\cite{brezin1976anomalous} that the anomalous dimension of a rank-$l$ tensor in $O(n)$ NL$\sigma$M is
\begin{equation}
    \Delta[l]=\frac{l(l+n-2)}{2}g+O(g^3)
    \label{eq:scalingtensor}
\end{equation}
up to higher order terms of $g$. Thus the dimension of rank two tensor is $\sim2.3$. 

We should also examine the large $k$ limit in $2d$. From Eq.\ref{eq:scalingtensor} we take the large $k$ limit and substitute in $g=g_c$, end up with $\Delta[2]=\frac{4}{k}$.
A symmetric traceless rank 2 tensor of $SO(4)$ corresponds to the $(1,1)$ operator in $SU(2)\otimes SU(2)$ CFT, with scaling dimension $\frac{4}{k+2}$ in $2d$, which is consistent with our result.

We can also estimate the drift in operator scaling dimensions. Assuming at system size $L_0$ the coupling constant reaches $g^*$. We consider system size $L$ that is not too far away from $L_0$, specifically $|\ln (L/L_0)|\ll \ln(\xi/\xi_0)$. Integrating the RG flow equation Eq.~\eqref{flowequation2} from $L_0$ to $L$ gives
\begin{equation}
    \delta g=-\frac{4\alpha}{9k^2}\ln (L/L_0)+...
\end{equation}
By Eq.~\eqref{eq:scalingtensor} we have (with $k=1$ and $\alpha=0.3$)
\begin{equation}
    \frac{\Delta_l(L)-\Delta_l(L_0)}{\Delta_l(L_0)}\approx-0.23\ln (L/L_0).
\end{equation}

\bibliography{SO(5)}

\end{document}